\pdfoutput=1
\documentclass[]{spie}
\usepackage{epsfig}
\usepackage[]{graphicx}
\usepackage{subfig}
\usepackage{soul}
\usepackage{color}
\usepackage{url} 
\usepackage{booktabs}
\usepackage{siunitx}
\usepackage[super]{nth}
\usepackage{amsmath}
\usepackage{enumitem}
\usepackage{cite}
\usepackage{float}

\newcommand{\degree}{$^{\circ}\,$}
\newcommand{\elec}{\mathrm{e^-}}

\title{Infrared tip-tilt sensing: on-sky experience, lessons learned and unsolved problems}

\author[a]{Marcos A. van Dam}
\author[b]{Sylvain Oberti}
\author[b]{Johann Kolb}
\author[c]{Jim Lyke}
\author[c]{Sylvain Cetre}
\author[d]{Beno\^it Neichel}

\affil[a]{Flat Wavefronts, 21 Lascelles Street, Christchurch 8022, New Zealand}
\affil[b]{European Southern Observatory, Karl-Schwarzschild-Str. 2, 85748 Garching, Germany}
\affil[c]{W.~M. Keck Observatory, 65-1120 Mamalahoa Highway, Kamuela, HI 96743, USA}
\affil[d]{Aix Marseille Universit\'e, CNRS, CNES, LAM, Marseille, France}

\authorinfo{Further author information: send correspondence to Marcos van Dam: marcos@flatwavefronts.com}

\begin{document}
\maketitle

\begin{abstract}
Infrared tip-tilt sensors (IR TTSs) have been deployed on three different laser guide star adaptive optics (AO) systems on three different telescopes. These IR TTS benefit from the high-order loop PSF sharpening in the near infrared, hence they provide a low tip-tilt residual and a good sky coverage. Nevertheless, these IR TTS are challenging and their use in AO is limited. In this paper, we outline existing IR TTS, provide on-sky performance results and describe our experience using IR TTS and along with plans for the near future. The second part of the paper deals with unresolved challenges for IR TTS. These include algorithms and loop stability in the low Strehl regime, using IR TTSs to measure higher-order modes and guiding on multiple guide stars with different magnitudes.

\end{abstract}

\section{Introduction}
\label{sec:intro}

Laser guide star adaptive optics (LGS AO) systems have been in routine operation on 8-10 \SI{}{\meter} class telescopes since 2004.\cite{LGSAOoverview} An LGS AO system measures the high-order turbulence with one or more wavefront sensors (WFSs) guiding on LGSs. The most common type of LGS is the sodium LGS, which uses laser light at a wavelength of \SI{589}{\nano\meter} to produce backscatter from sodium atoms in the mesosphere at an altitude between \SI{80}{} to \SI{100}{\kilo\meter} above sea level. 

In addition, a natural guide star (NGS) is needed to make fast tip-tilt measurements, because the tip-tilt measurements from the LGS are contaminated by turbulence on the up-link path. For sodium LGSs, the WFS sees an elongated star with a time-varying focus due to the changing sodium density and altitude. The NGS used for tip-tilt sensing is typically also used to measure these changes in focus, and sometimes to also measure aberrations induced by the elongated LGS seen by the WFS.\cite{LGSAOoverview} 

Until very recently, the tip-tilt measurements on all LGS AO systems in routine science operation were made using visible, and not infrared (IR), light. There are three reasons for this. First, AO systems typically use the infrared wavelengths for science. Second, infrared cameras are much more expensive than their visible light counterparts. Third, until recently, IR detectors operating in the \SI{100} to \SI{1000}{\Hz} frame rate range suffered from high read noise. Despite these issues, IR TTSs can be very useful for LGS AO systems on 8-10 \SI{}{\meter} class telescopes and indispensable on the next generation of extremely large telescopes (ELTs) with diameters ranging between \SI{25}{\meter} and \SI{39}{\meter}. Recall that the error in estimating the displacement of a spot is inversely proportional to the size of the spot. The high-order loop of an LGS AO system sharpens the point-spread function (PSF) and produces diffraction-limited cores in the IR. These PSFs have a resolution of the order of $\lambda/D$, where $\lambda$ is the wavelength and $D$ is the diameter of the telescope. For 8-10 \SI{}{\meter} class telescopes operating at K-band (\SI{2.2}{\micro\meter}), images at the diffraction limit have a FWHM of the order of \SI{50} milliarcseconds (mas). This is already 10 times smaller than a visible light seeing-limited spot, even in very good seeing! The benefits increase with increasing telescope diameter and decreasing wavelength, provided that the diffraction limit can be maintained. Hence, the designs of all of the AO systems for the ELTs include IR TTSs. 

In spite of the fact that IR TTSs will be ubiquitous in the near future, astronomers and instrumention specialists have very limited experience with their use. There are currently three different telescopes utilizing three different flavors of AO with IR TTSs installed, and each of these systems is described in detail in Sections \ref{sec:GeMS}, \ref{sec:Keck} and \ref{sec:VLT} respectively. The Gemini South multi-conjugate adaptive optics (MCAO) system, GeMS, has the ability to guide on three on-detector guide windows (ODGWs) on the science camera. The ``classical'' LGS AO system at the W.~M. Keck Observatory has a dedicated IR TTS that picks up the light from one or more stars over a large field of regard just before the science instrument. The most recent arrival is infrared low-order sensor (IRLOS) for the laser tomography adaptive optics (LTAO) system at the VLT, which uses the science target or a star within 5 arcsec of the optical axis to correct tip-tilt. Since April 2019, it has been offered to astronomers for regular operations.

All of these systems Teledyne Hawaii-xRG HgCdTe detectors, which consist of $1000\times1000$ or more pixels. The process of reading the whole frame takes too long and is too noisy to be used for tip-tilt sensing. However, ODGWs can be read at a faster frame rate and with a lower read-out noise. Recently, HgCdTe IR electron avalanche photo-diode (eAPD) arrays have become commercially available\cite{SAPHIRA} and are being used for adaptive optics applications.\cite{NIRWFSWizinowich,GMTDFS} These devices have the following advantages with respect with Hawaii-xRG detectors:
\begin{itemize}[nolistsep]
\item[--] Much lower cost
\item[--] Outstanding cosmetics (very few bad pixels)
\item[--] High full-frame frame rate
\item[--] Sub-electron read noise with only a modest excess noise factor
\end{itemize}
ESO has begun a project to update their IR TTS to use a SAPHIRA array in LTAO mode in order to increase the limiting magnitude by two to three magnitudes. On the other hand, these eAPD arrays have a smaller number of pixels (currently $320\times240$) and the field-of-regard would be limited unless they are placed on a translating stage. 

The purpose of this paper is to collate the experience of the AO scientists on these three instruments in the hope that it is useful in the design and implementation of future systems. Remaining issues which need to be addressed by for future systems are described in Section \ref{sec:openquestions}. Finally, conclusions are drawn in Section \ref{sec:conclusion}.

\section{Infrared tip-tilt sensing at Gemini South}
\label{sec:GeMS}

\subsection{Overview}
The Gemini South MCAO system, known as GeMS, is the first and only LGS-based MCAO system in operation.\cite{GeMSI,GeMSII} High-order measurements are made by five Shack-Hartmann wavefront sensors (SH WFSs) with $16\times16$ subapertures across the pupil, guiding on a fixed LGS asterism that ressembles the number five on a dice. There are two deformable mirrors (DMs): a DM conjugate to the ground with 240 actuators in the pupil, and one at a conjugate altitude of \SI{9}{\kilo\meter} with 120 actuators. Tip-tilt is controlled by a dedicated tip-tilt mirror guiding on up to three visible tip-tilt stars, and quadratic modes (focus and astigmatism) are injected into the high-altitude DM in order to correct the so-called plate scale modes. An MCAO system requires three NGSs to control the global and field-dependent image shifts induced by the atmosphere and unsensed by the LGSs.

During the design of GeMS, two modes of operations were foreseen to measure these modes. The baseline operation mode for GeMS is to use three visible-light probes spanning a field-of-regard of 2 arcminutes. Each probe comprises of a custom arrangement of APDs and one of the probes also has an associated focus sensor. The whole assembly which will be replaced in 2019 by a new tip-tilt sensor that comprises of a single $512\times512$ pixel CCD with electron multiplication to reduce the effect of the read noise.\cite{NGS2} However, in some very embedded fields ({\em e.g.}, the Orion Nebula and the Antenna Galaxy), the probability of finding a suitable star at visible wavelengths is too low. The second mode of operation for tip-tilt control makes use of on-detector guide windows (ODGWs) in the infrared imager called GSAOI, which has an array of $2\times2$ Teledyne Hawaii-2RG detectors with a total of $4080\times4080$ 20 mas pixels.\cite{Young} Each of the four detectors can read one ODGW. The ODGWs are used in a fast read-out mode, providing tip-tilt information based on the centroid position of the stars. In this mode, one visible star (one of the three probes) is still needed for slow focus compensation. 

\subsection{Operations}

\begin{figure}[htbp]
  \begin{center}
 \includegraphics[width=\linewidth]{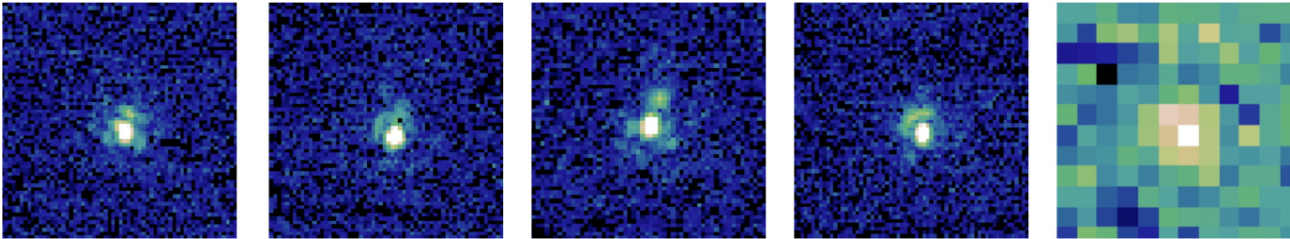}  
  \caption{GSAOI ODGW images of tip-tilt stars: the first four are \SI{10}{\milli\second} exposures, while the fifth is a \SI{1.6}{\milli\second} exposure.\cite{Young}}
  \label{fig:gsaoiodgw}
   \end{center}
\end{figure}

Tests of tip-tilt control using the ODGWs were done during the commissioning of the instrument, and were successful. Fig. \ref{fig:gsaoiodgw} shows five on-sky K-short short exposure images obtained using the ODGWs that were used to close the tip-tilt loop.  
However, there were instabilities in the detector controller that caused the GSAOI detector controller to crash when a full detector frame was read out while guiding on the ODGWs. There are no plans in the pipeline fix this problem and use the ODGWs for tip-tilt sensing. The ODGWs are still used for slow guiding, compensating for differential flexures between the AO bench and the science camera. The acquisition procedure for flexure compensation is described below.

The ODGW position is determined based on a mapping between telescope focal plane and instrument focal place. The size of the ODGW is set to its largest value, $128\times128$ pixels, or $2.5\times2.5$ arcseconds on-sky. If the star does not appear on the ODGW, a full-frame image can be taken with GSAOI to check its position with respect to the guide window. The stars are properly centered on the ODGW by re-defining the ODGW $x$ and $y$ detector position and not by moving the telescope, since all of the loops are closed at this point. Once the star is properly centered on the ODGW, the size of the ODGW window is reduced to $16\times16$ pixels and the flexure loop is closed. 

The main challenge of sharing the same focal plane array for both science and guiding resides in the fact that different filters may have different settings. If the ODGW is used in a sequence that requires several filters, the exposure time of the ODGW must be adjusted to depending on the SNR. On the other hand, having the ability to measure the PSF at the science focal plane was found to be an extremely useful diagnostic tool during commissioning, and could be envisioned as an operational performance diagnosis tool. Indeed, one has a direct access to short exposure PSFs, and AO loop parameters may be optimized based on the measured image quality. It was used to demonstrate that the vibrations seen at the instrument focal plane were common to both the science instrument and the WFSs. The ODGWs were also to run a fast scan through focus and determine the zero-point of the focus sensor. The use of ODGWs to monitor dynamic focus shifts due to the changing sodium altitude is described in Section \ref{sec:lift}.
 
\section{Infrared tip-tilt sensing at the W.~M. Keck Observatory}
\label{sec:Keck}

\subsection{Overview}
The two telescopes at the W.~M. Keck Observatory are equipped with almost identical AO systems, with high-order correction using a 349-actuator Xinetics DM based on a $20\times20$ SH WFS guiding on either an NGS or an LGS.\cite{LGSAOoverview,LGSAOperformance,KeckAOUpgrade} The tip-tilt sensor used in LGS AO mode is STRAP, a $2\times2$ array of avalanche photodiodes using visible light, which drives a custom tip-tilt mirror. 

The Keck I AO system is additionally equipped with an IR TTS, TRICK, described in detail elsewhere.\cite{TRICK2014,TRICK2015,TRICK2016} H- or K-band light is picked off with a dichroic beamsplitter in front of OSIRIS, a near infrared spectrograph and imager, and sent to TRICK. The TRICK detector is an engineering grade Teledyne Hawaii-2RG detector, with $2048\times2048$ 50 mas pixels for a total field-of-regard of $102\times102$ arcseconds. The detector and reimaging optics are cryogenically cooled to \SI{105}{\kelvin} to maintain low noise up to \SI{2.2}{\micro\meter}. TRICK can read up to eight regions of interest (ROIs), with each ROI having $2\times2$, $4\times4$, $8\times8$ or $16\times16$ pixels. The frame rate can be as high as \SI{8}{\kilo\Hz}, although the wavefront controller can only accept commands at \SI{2}{\kilo\Hz}.

The full detector frame can also be read in order to find the tip-tilt stars in the field. This operation takes about \SI{20}{\second}.

\subsection{Operations}

A very simplified description of the current acquisition sequence for the use of the Keck I LGS AO in conjunction with TRICK is described below. 
A visible light tip-tilt star is still required, both for acquisition and to measure focus and LGS aberrations. This is usually, but not necessarily, the same as the infrared tip-tilt star. 
The original concept of acquiring directly on the IR TTS was abandoned as on-sky experience showed that it was too difficult to center the star on the ROI before closing the tip-tilt loop. The acquisition steps are as follows: 
\begin{enumerate}[nolistsep]
\item Acquire a tip-tilt star on STRAP on the pointing origin of the science instrument and close the tip-tilt loop.
\item Acquire the LGS, close the up-link tip-tilt loop, then close the DM loop on the LGS.
\item Measure focus errors and LGS aberrations with the visible light star on the low-bandwidth WFS.
\item Open the loops, offset the telescope to the science target and reclose the loops.
\item Set up a $16\times16$ ROI centered on the expected location of the IR tip-tilt star. If the star is not in the ROI then
\begin{itemize}[nolistsep]
\item[--] Move the ROI is a spiral pattern until you find the star, or
\item[--] Take a full detector frame and select a star in the frame.
\end{itemize}
\item If necessary, adjust the coordinates of the ROI to center the tip-tilt star.
\item Reconfigure the ROI to $4\times4$ pixels.
\item Take a sky background in a nearby ROI.
\item Switch control of the tip-tilt mirror from STRAP to TRICK.
\end{enumerate}
The switch from STRAP to TRICK takes about \SI{30}{\second} if the star is initially within the $16\times16$ acquisition ROI. 

\begin{figure}[htb]
  \begin{center}
 \includegraphics[height=10cm]{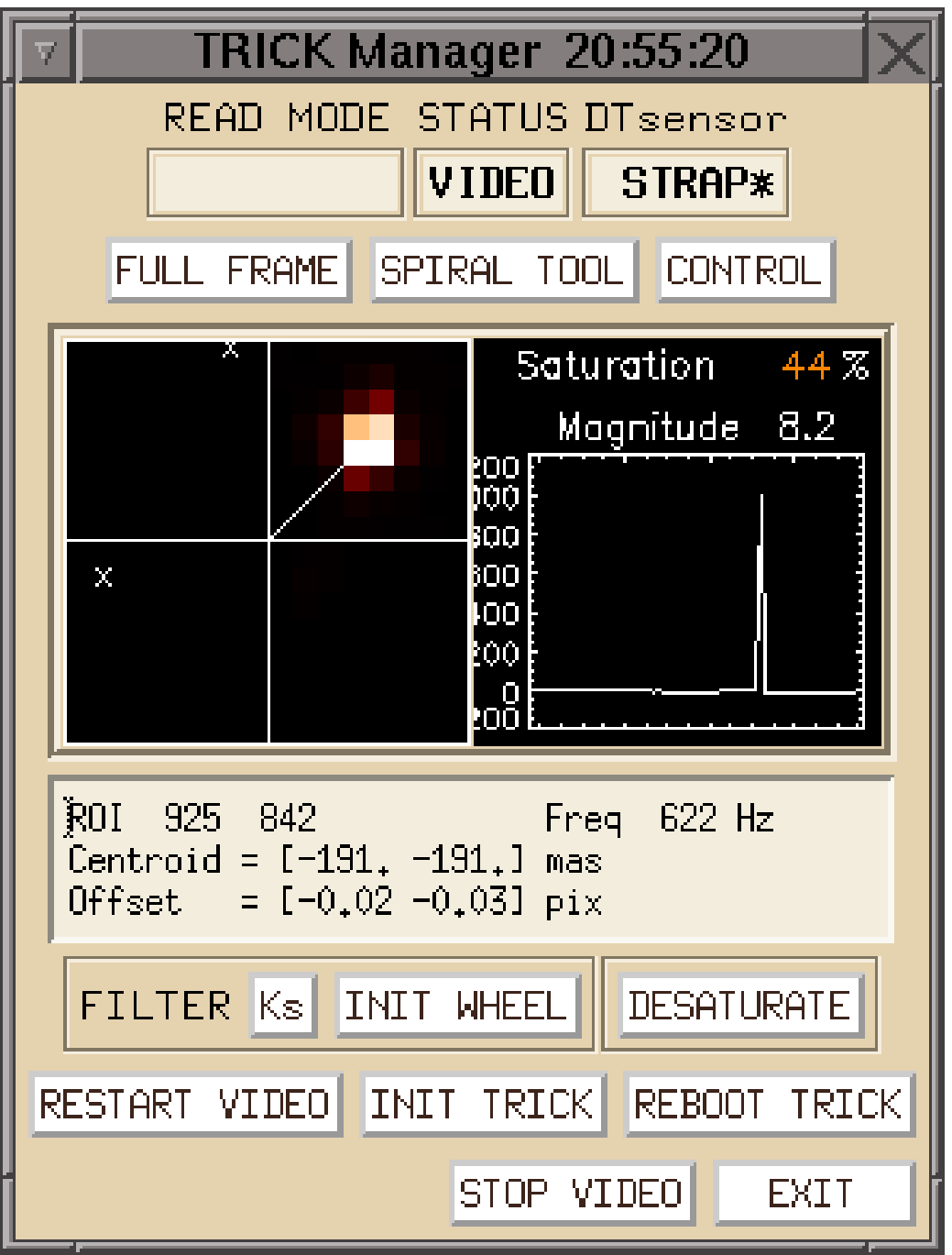}  
 \includegraphics[height=10cm]{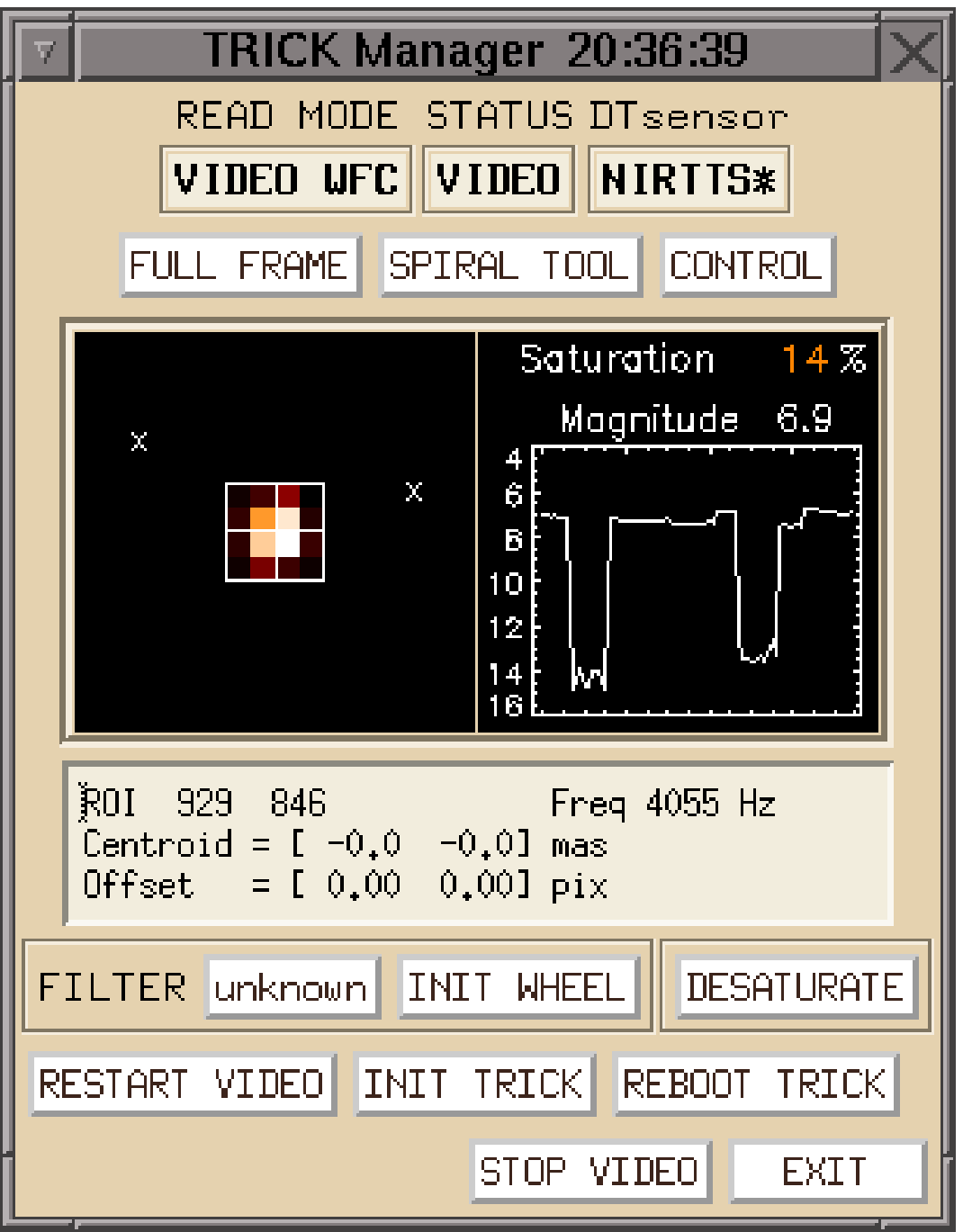}  
 \caption{TRICK acqusition tool when acquiring a star with a $16\times16$ pixel ROI (left) and after the tip-tilt loop is closed on a $4\times4$ pixel ROI (right). By clicking on the tip-tilt star, the ROIs are redefined so that the star is at the center. The $\times$ symbol represents bad pixels on the detector.}
  \label{fig:trick_acquisition}
    \end{center}
\end{figure}

Fig. \ref{fig:trick_acquisition} shows the GUI used to close the tuip-tilt loop on TRICK.
The engineering grade detector has a large number of bad pixels, denoted by a $\times$ symbol on the GUI, and these pixels must be avoided for optimal performance. The distance between the guide star and the science target are fixed, but the rotator angle on the AO bench can be selected to avoid regions of bad pixels. 

When telescope offsets are requested by the astronomer, the loops open, the telescope moves, the up-link tip-tilt and DM loops close, and then the tip-tilt loop closes. Alternatively, we can reclose the tip-tilt loop on STRAP before handing over control of the tip-tilt mirror to TRICK.

Since the science wavelength and the TTS wavelengths differ, changes in telescope elevation lead to differential atmospheric refraction (DAR). This is compensated by changing the centroid origins. When the value of the centroid origins exceed half a pixel, the ROI should be redefined, but this function has not yet been implemented. 

The system has proven to be stable on the sky for easy observations, where the tip-tilt star is relatively bright ($m_H \le 15$) and near on-axis ($\theta \le$ 30 arcsec). On the other hand, for guide stars $\ge$ 30 arcsec off-axis, the time-variable PSF can lead to loss of tip-tilt control. This problem could be mitigated by using larger ROIs; however, this implies the use of more pixels with a higher read noise per pixel, so the centroiding error rapidly increases. The effect of the measurement noise increasing with increasing number of pixels could be bypassed through the use of the correlation algorithm,\cite{LGSAOperformance} which has been implemented but does not currently work as intended.

\subsection{Performance}
An outstanding feature of the TRICK detector readout is the very low noise. The detector is continuously operated in a non-destructive readout sequence. The readout time is \SI{44}{\micro\second} for $2\times2$ pixels and \SI{123}{\micro\second} for $4\times4$ pixels. Low read noise is achieved by non-destructively reading the same ROI multiple times before coadding the result. Increasing the number of coadds reduces the effective read noise but reduces the tip-tilt correction bandwidth. For a $4\times4$ pixel ROI, the read noise per pixel is $\le 3\elec$ for frame rates between \SI{30}{\Hz} and \SI{200}{\Hz}, rising to $4\elec$ and $5\elec$ for frame rates of \SI{600}{\Hz} and \SI{1000}{\Hz} respectively.\cite{TRICK2014,TRICK2016,Smith} The maximum frame rate is \SI{8000}{\Hz}, although neither the wavefront controller nor the tip-tilt mirror can accommodate this bandwidth.

\begin{figure}[htb]
  \begin{center}
 \includegraphics[width=10cm]{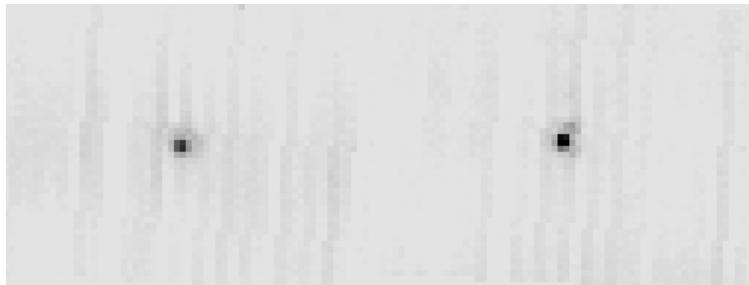}
 \caption{OSIRIS spectrograph H-band image using STRAP (left) and TRICK (right) while guiding on a $m_R=13$ star.}
  \label{fig:trickbrightpsf}
    \end{center}
\end{figure}

Fig. \ref{fig:trickbrightpsf} shows a comparison of the science PSF delivered using a visible light TTS and an IR TTS. Switching from STRAP to TRICK led to a 50\% increase in relative H-band Strehl and to a reduction in FWHM from 46.8 mas to 37.8 mas.
With the tip-tilt loop closed at\SI{2}{\kilo\hertz} on TRICK, the tip-tilt bandwidth error was kept below \SI{50}{\nano\meter}.

Recent work has focused on operability and reliability, rather than performance. The reader is referred to previous work by Rampy {\em et al.}\cite{TRICK2015} for performance results.

\subsection{Future work}

TRICK has not yet been offered to the scientific community. The integration of the hardware and user tools into an already operational system has caused many headaches,\cite{TRICK2015,TRICK2016} but almost all of the issues have been overcome by rewriting the user tools and making changes to the operational model. The last remaining significant issue is that the AO server becomes overloaded when TRICK is operating, so the server is in the process of being replaced. A small and decreasing number of software bugs remain,  while some functionality remains to be added:
\begin{itemize}[nolistsep]
\item[--] Automatic transition of tip-tilt control to STRAP when DM loops are open ({\em e.g.}, if the laser is shuttered)
\item[--] DAR correction by changing ROI on the fly during an observation  
\item[--] Automation of loop optimization
\item[--] Detection of and recovery from loss of tip-tilt control
\item[--] Implementation of the correlation algorithm to find spot displacement
\end{itemize}
The subsequent step is to run a range of science verification cases, and then offer it to the community in shared-risk mode. 

TRICK has a very bright future because funding has been obtained to upgrade the Keck I AO system into an LTAO system called Keck All-Sky Precision Adaptive Optics (KAPA). The LTAO design calls for the use of two or three tip-tilt stars in order to reduce the effect of tip-tilt anisoplanatism. The capability to use multiple ROIs is already present; how to optimally combine the measurements is work in progress, and this is discussed in Section \ref{sec:multiplegs}. Currently, the measurements from multiple stars can only be combined using a weighted average,\cite{TRICKmultipleGS} and an upgrade to real-time controller code is needed to support tomography.

\section{Infrared tip-tilt sensing at the Very Large Telescope}
\label{sec:VLT}

\subsection{Overview}

The GALACSI Narrow Field Mode (NFM) is the AO module offering Laser Tomography AO correction to MUSE,\cite{obertispie2018} and uses IRLOS as a tip-tilt and focus sensor. IRLOS consists of a $2\times2$ SH WFS operating in J+H band, based on an Hawaii-1 detector. In addition to correcting the fast atmospheric and wind-shake tip-tilt, IRLOS is used as a slow truth sensor to track the defocus due to changing sodium altitude and, optionally, the two astigmatism modes.

IRLOS has two pixel scales: 60 mas used when guiding on unresolved objects and 250 mas used when guiding on extended targets. As a comparison, the Nyquist sampling pixel scale 32 mas for a \SI{4}{\meter} subaperture at a wavelength of \SI{1.25}{\micro\meter}. Centroid offsets are applied to ensure that the IRLOS focus agrees with the focus of the science instrument. 

\subsection{Operations}

A simplified description of the current acquisition sequence for the use of the MUSE/GALACSI narrowfield mode (NFM) at the VLT is described below, with an emphasis on the steps relating to the IR TTS. 

\begin{figure}[htbp]
  \begin{center}
 \includegraphics[]{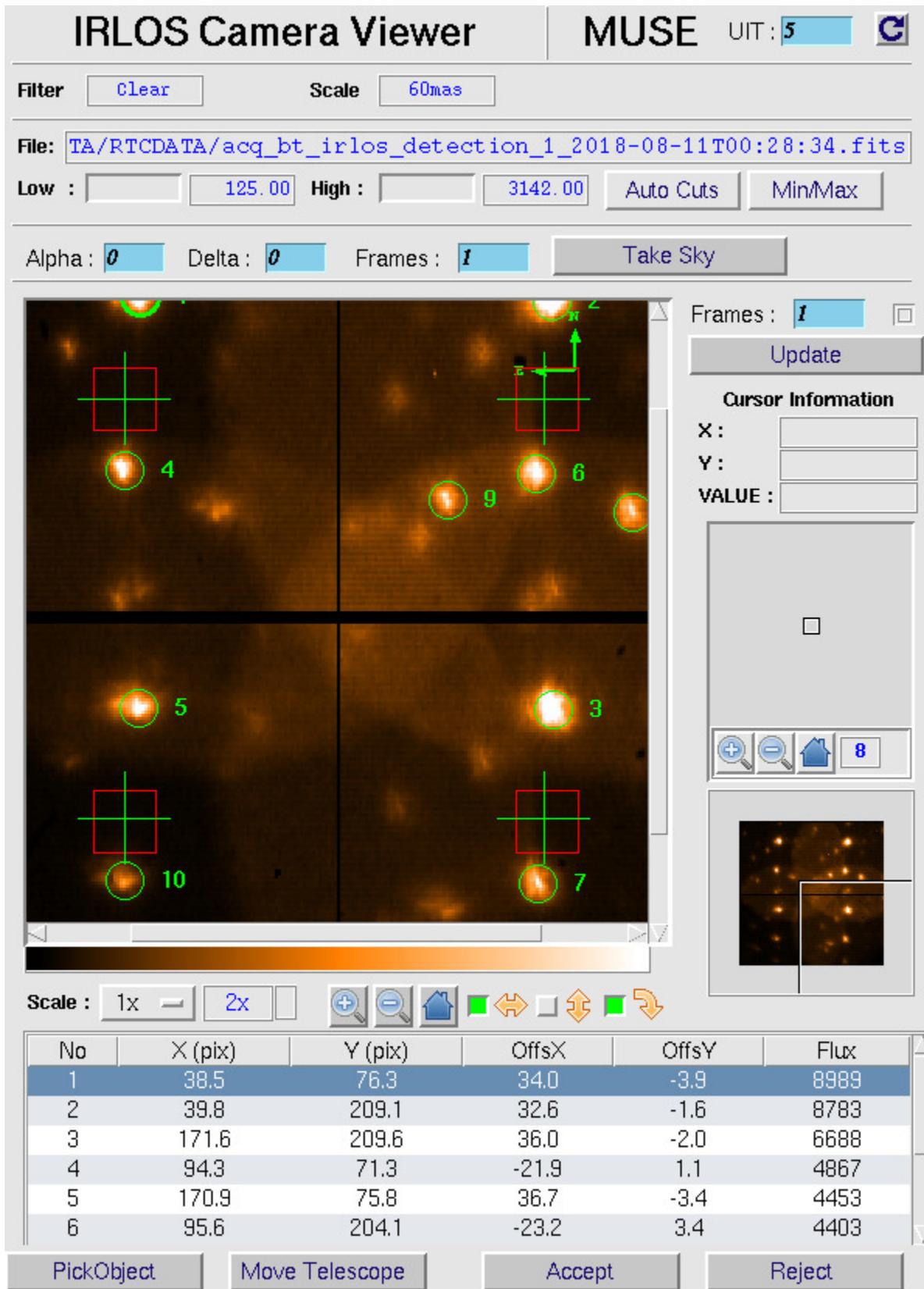}  
  \caption{Acquisition tool used to acquire the infrared tip-tilt object on the InfraRed Low-Order Sensor (IRLOS). An image of the $2\times2$ sub-apertures is shown, with overlapping subimages due to an oversized field stop.}
  \label{fig:irlos_acquisition}
    \end{center}
\end{figure}

While slewing to a new target, all the motors are set to use AO and the real-time controller (RTC) is switched to the NFM configuration. Once the telescope is guiding and has performed one Active Optics correction (to get a decent shape on M1), the LGSs propagated and the pointing set using look-up tables. If the guide stars are not visible on the four WFS, we search for them in a $12\times12$ arcsecond square FoV by actuating the fast jitter mirror on the launch telescopes. The search takes $4\times10$\SI{}{\second} (the search has to be done sequentially as the range is larger than the LGSs separation so there is overlap on the other WFSs) and ensures that the LGSs are acquired 100\% of the time. The next phase consists in closing the high-order loop which sharpens the IR guide object before it is acquired. Focus is first minimized on the LGS WFSs using the Focus Compensating Device, as optic that changes the distance to the WFSs in order to minimize the focus error. Next, a precomputed system control matrix is loaded, which accounts for the DM/WFS mis-registrations and the distance to the LGS at the current telescope elevation. GALACSI has no pupil centering mirrors. The telescope Active Optics are stopped and the loops closed between a) the LGS WFSs jitter and fast actuators in front of the WFSs, and b) the LGS WFSs high orders and the Deformable Secondary Mirror. An image is then taken with IRLOS in its acquisition configuration: $140\times140$ 60-mas pixels per subaperture, \SI{1}{\second} exposure, as shown on Fig. \ref{fig:irlos_acquisition}. 
The desired object is manually selected in any of the four sub-apertures and the telescope is offset to put it at the center of the respective region-of-interest (ROI) window. Meanwhile, the telescope is stabilizing the field at \SI{65}{\hertz} so that the image position is stable. Finally the detector readout rate is increased to \SI{207}{\hertz}) around the ROIs and several steps ensure that the star is still in the the $8\times8$ pixel ROI before closing the TT loop:
\begin{itemize}[nolistsep]
\item[--] If the star is faint, a dark is measured.
\item[--] The focus on IRLOS iteratively minimized by sending offsets to the LGSs Focus Compensating Device.
\item[--] Switch at the last possible moment the control of the tilt from the telescope to the AO system. 
\item[--] Engage the sky follower loop that changes the detector background to minimize its value at the edge of the sub-apertures, and after three cycles,
\item[--] Prompt the user to check for the presence and centering of the star before closing the TT loop.
\end{itemize}
Following this process, secondary loops are engaged to adapt the weighting map shape to the spot size, numerically derotate the IR TT control matrix and send offsets to the LGSs Focus Compensating Device from averaged focus residual measurements on the IR WFS. The procedure is fully automated except for the selection of the IR star and the confirmation that it is centered before closing the loop. It takes in total between \SI{6.5}{\minute} and \SI{10.5}{\minute}, including the telescope slewing time. Software improvements have been proposed to reduce this time by at least a minute.

During an observation sequence, small offsets (dithering) and derotator offsets are allowed while the loops are closed, and larger offsets and sky measurements are applied by opening, counter-offsetting IRLOS and re-closing the TT loop.

During an exposure, if an airplane detection interrupts the LGS propagation (an event which happens about twice a night and lasts about \SI{10}{\second}) the high-order loop freezes and the IR star image becomes uncorrected, leading to unstability. At that moment the MUSE exposure is paused using a fast mechanical shutter until the correction is resumed. If the TT star is lost (e.g., due to a pause in LGS propagation, a burst of bad seeing or when guiding on a very faint TT star), the TT star is reacquired in less than \SI{2}{\minute} at the push of a button.

\begin{figure}[htbp]
  \begin{center}
 \includegraphics[width=0.3\linewidth]{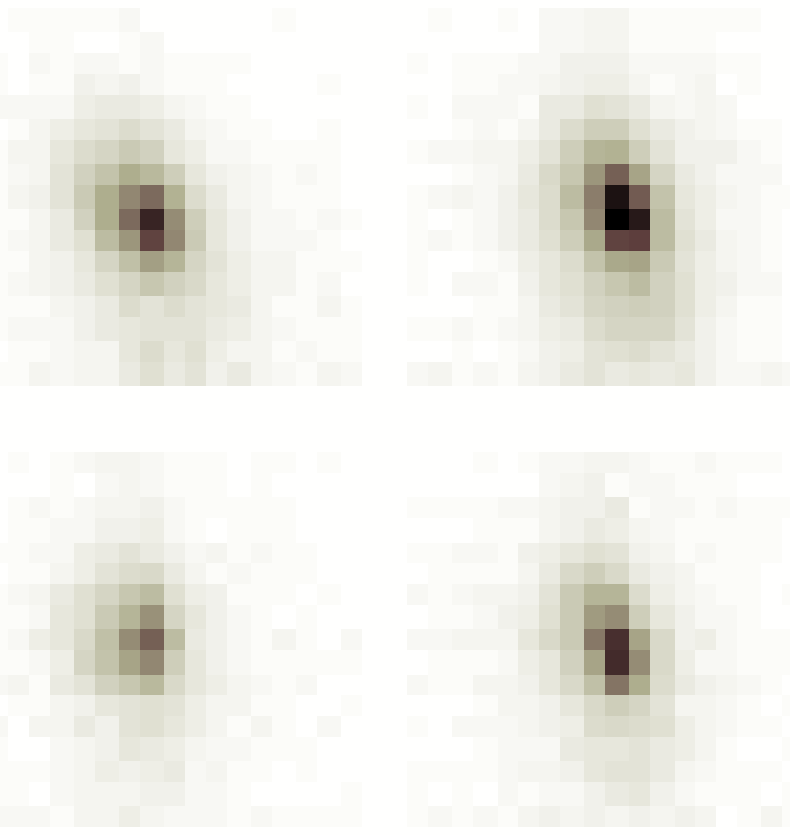}  
 \includegraphics[width=0.3\linewidth]{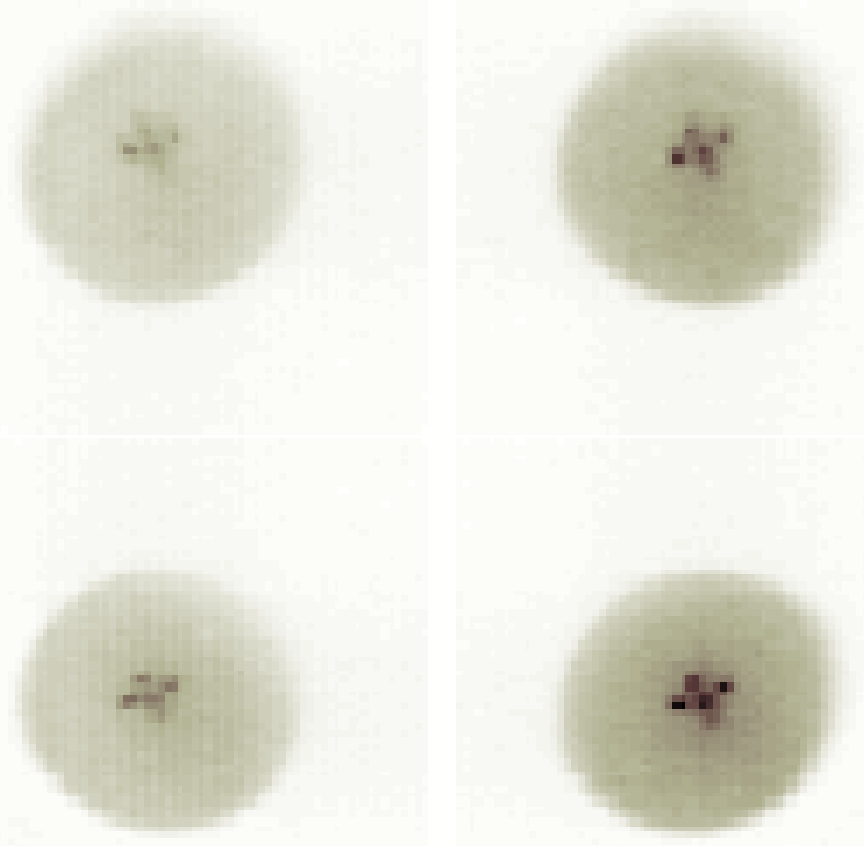}  
  \caption{\SI{1}{\second} exposure acquisition images of NGC 6810 (left) and Einstein's Cross (right).}
  \label{fig:ngc6810}
   \end{center}
\end{figure}

IRLOS can acquire and close the loop on extended targets of up to 2 arcseconds in diameter, like cores of galaxies or quasars. Fig. \ref{fig:ngc6810} shows the four subimages of NGC 6810, a spiral galaxy which was suitable as a TT object, and the Einstein Cross, a gravitationally lensed quasar, which was too faint to guide on at \SI{207}{\hertz}.
The frame rate can be increased to \SI{500}{\hertz} on bright targets by reducing the FoV to $6\times6$. This mode has not been commissioned for routine science operation due to lack of time.

When the observation is finished, all the loops are opened in the right order (offload loops, TT, high-order control and jitter control) the laser propagation is stopped and full control is given back to the telescope. 

\subsection{Performance}

The performance of the NFM using IRLOS was estimated using the commissioning camera. The bright tip-tilt star performance is outstanding: the telescope diffraction limit of 25 mas at a wavelength of \SI{950}{\nano\meter} is attained, with a Strehl ratio of 30\%. At \SI{650}{\nano\meter}, the FWHM is still 25 mas, close to the diffraction limit of 17 mas.

\begin{figure}[htbp]
  \begin{center}
 \includegraphics[width=1\linewidth]{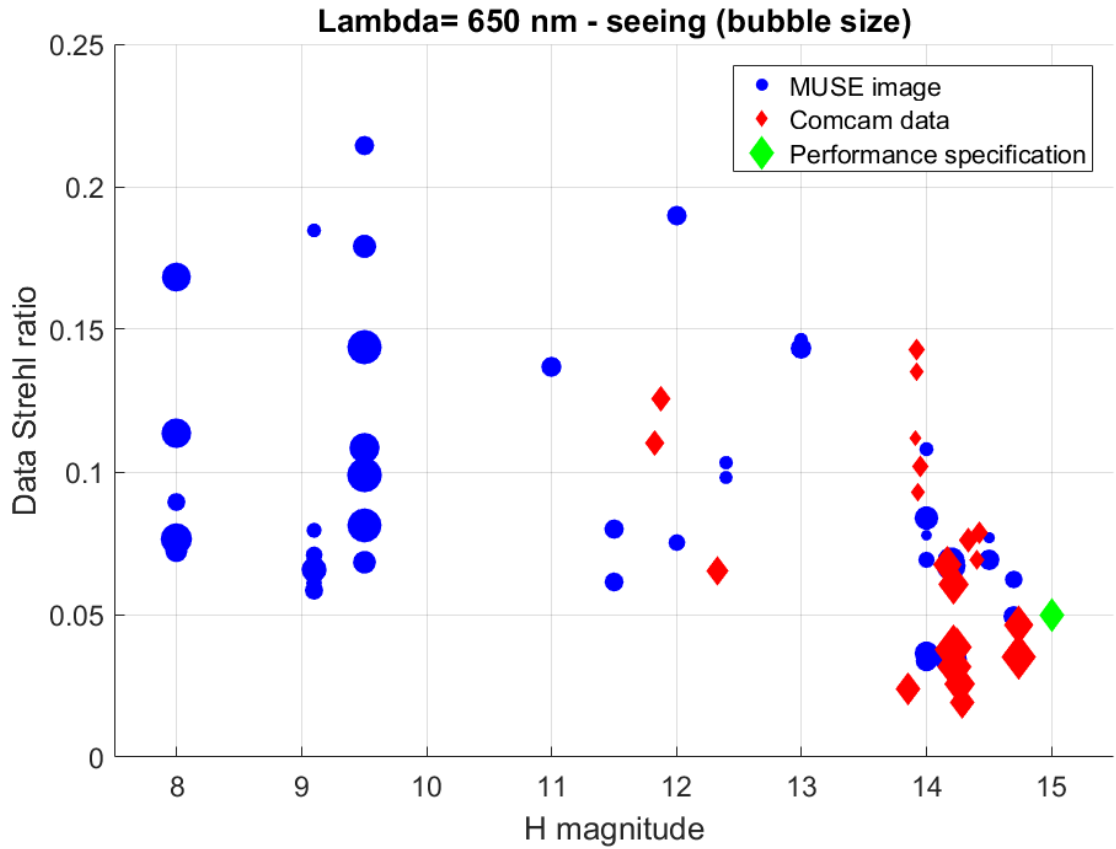}  
 \caption{Strehl ratio measured at \SI{650}{\nano\meter} versus H-band magnitude, with the bubble size encoding the seeing along the line of sight. The blue circles represent MUSE images, while the red diamonds are images taken with the commissioning camera. The data was taken vefore the realignment mission in November 2018.}
  \label{fig:irlosperf}
   \end{center}
\end{figure}

Originally IRLOS was misaligned in pupil and focus, creating un unbalance of flux between the four subapertures and larger spots than expected. The performance with the misalignment in place is shown in Fig. \ref{fig:irlosperf}. An intervention made in November 2018, before the system was offered to the science community, has improved the limiting magnitude.
Even on bright stars, the error budget is dominated by tip-tilt bandwidth errors, since the only mode offered is $8\times8$ pixels at a frame rate of \SI{200}{\hertz}. The detector can read a $6\times6$ window at \SI{500}{\hertz}. The performance drops off for stars fainter than $m_H=12$ due to photon noise, and the measurement error is dominated by the read noise of and read noise of the detector, which is about $11\elec$, for stars $m_H\ge 14$. Stars down to $m_H=15$ are attainable in excellent seeing. The low frame rate and high read noise will be addressed through an upgrade described in Section \ref{sec:irlosfuture}.

The IRLOS performance is affected by differential piston caused by the so-called low wind effect, where the temperature differential on either side of the telescope spiders results in a piston difference and leads to a breakdown in the PSF core. A reflective foil has been applied to the telescope spider in order to reduce the temperature difference.

Vibrations in the tip-tilt and defocus modes are tracked and cancelled. However, this strategy has not been successful because the main vibration at the VLT is trefoil, which aliases onto tip-tilt and focus when measured with a $2\times2$ SH WFS. Instead, the trefoil vibration is reduced by shaping the response of the high-order loop temporal controller in order to cancel trefoil vibrations at \SI{45}{\hertz}.

\subsection{Future work}
\label{sec:irlosfuture}

In this section, we describe work in progress to improve the performance of IRLOS.

The performance of an IR TTS depends on the PSF delivered by the high-order loop. The photon noise variance is inversely proportional to the Strehl ratio while the detector noise (read-out noise and background) is inversely proportional to the Strehl ratio squared. A mission is planned in October 2019 to optimize the LTAO reconstructor in two ways:
\begin{itemize}[nolistsep]
\item[--] Improve the model geometry and the number of reconstructed layers to reduce the tomographic error.
\item[--] Improve the misregistration model and increase the number of corrected modes to minimize the fitting error term.
\end{itemize}

A detector upgrade project is under way to replace the Hawaii-1 detector by a SAPHIRA e-APD array.\cite{SAPHIRA} The sub-electron read noise will reduce the measurement noise error and extend the limiting magnitude by two to thre magnitudes. The use of a SAPHIRA array will permit read outs at a frame rate of \SI{500}{\hertz} for most targets, reducing the temporal bandwidth error.

It is also essential to reduce the background noise, dominated by the sky background at wavelengths less than \SI{1.8}{\micro\meter} and the thermal background at longer wavelengths. The thermal background must be minimized by means of a cold IR filter, as the SAPHIRA detector is sensitive up to \SI{3.7}{\micro\meter}. For large APD gains, filter of optical density of four is sufficient, while at small APD gains, an optical density of five or higher is required. This is assumes that the thermal baffling has been properly managed with cold baffles at different conjugated planes to reduce the parasitic stray light coming from emissive objects outside the field of view of interest. These multiple baffles are also important to minimize the angle of incidence on the interferential blocking filters, whose transmission is otherwise blue shifted.\cite{Gach} Finally, the thermal background shall be minimized by optical design, essentially by minimizing the optical etendue of the warm optics. This can be achieved by increasing the f-number and by extending the distance between the cryostat window and the detector plane. A cold imaging lens is also an option which was not considered for the IRLOS upgrade.

In order to increase further the limiting magnitude, a tip-tilt sensor using the full telescope pupil is proposed, thereby reducing the size of the PSF by a factor of two. A pixel scale of 18 mas samples the H-band PSF core very well. However with this scheme, the focus measuring functionality of a $2\times2$ SH WFS is lost and the sodium layer altitude variation cannot be tracked any longer. An attractive idea to solve this is to use the linearized focal-plane technique (LIFT) approach,\cite{LIFT,LIFT2016} discussed in Section \ref{sec:lift}. GALACSI NFM is the perfect case to apply LIFT, as the H-band correction is excellent, with a Strehl ratio in excess of 50\%, and the 18 mas pixels provide better than Nyquist sampling. The relationship between focus and the signal provided by LIFT is thus linear.

\section{Open issues for near infrared tip-tilt sensors}
\label{sec:openquestions}
This section addresses issues which must be considered in the design of the next generation of IR TTSs.

\subsection{Strehl ratio and sensitivity}
\label{sec:sensitivity}
IR TTSs are sensitive to residual wavefront aberrations. For a diffraction-limited core, the sensitivity of the sensor is inversely proportional to the Strehl ratio. We have to optimize the performance not only on the science target, but on the NIR TTS as well.

Static wavefront aberrations on the science path are typically measured during calibration and compensated by changing the shape of the DM. Provided that the response of the WFS is sufficiently linear and well modeled, these aberrations do not have an impact on the wavefront delivered by the AO system. However, any non-common path aberration between the science detector and the TTS will lead to wavefront errors at the TTS. To mitigate this, instruments under design include the TTS inside the science instrument. In addition, it may be necessary to to incorporate a DM in the path of the TTS, especially if we need to correct for anisoplanatism. Work is under way to chacterize the behavior of existing DMs in a cryogenic environment.\cite{EnvironmentalTestingDMs} Alternatively, we can compensate the whole field using MCAO, and this solution was preferred for the NFIRAOS instrument at the TMT. 

\begin{figure}[htbp]
\centering
\includegraphics[width=0.18\linewidth]{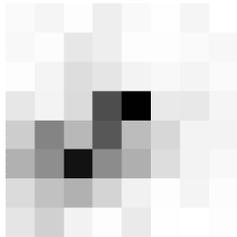}
\caption{Simulated K-band image for TRICK when guiding on an off-axis star. There are two peaks within the $8\times8$ pixel region, which presents a problem for a peak finding algorithm.}
\label{fig:offaxisspot}
\end{figure}

Traditional SH WFSs and tip-tilt sensors assume Gaussian spots ({\em e.g.}, centroid algorithm) or spots with a known or assumed shape ({\em e.g.}, correlation algorithm or matched filter). These approaches work well for diffraction-limited spots, seeing-limited spots, elongated spots from LGSs, and solar granulation (where the reference can be measured). Matters become complicated at modest Strehl ratios (below 10\%), where the core can degenerate into two or more speckles (Fig. \ref{fig:offaxisspot}). The spot on the TTS becomes larger due to anisoplanatism, so a larger detector region is needed. Since reading more pixels increases the read noise on each pixel, and since the error in the centroid measurement increases with number of pixels even if the noise were the same, then it follows that the centroid algorithm will suffer from increased noise. However, peak finding algorithms, such as the correlation algorithm, will track the brightest speckle, which will change with time and lead to an inaccurate tip-tilt estimate. Optimal spot displacement algorithms for this r\'egime are an open research problem.

A related issue is optimal tip-tilt sensing with extended objects, for which conventional centroiding algorithms are not well suited.  

\subsection{Tip-tilt sensing with multiple stars} 
\label{sec:multiplegs}
The problem of combining multiple, bright off-axis tip-tilt guide stars to obtain an on-axis tip-tilt estimate has been studied in the past, initially in the context of MCAO. Flicker presented a linear model for the change in tip-tilt across the field as a result of the three high-altitude focus and astigmatism (the so-called ``plate-scale'' modes).\cite{FlickerTTaniso} By making three off-axis tip-tilt measurements, we can estimate five terms (on-axis tip-tilt and the plate scale modes). 

A better solution for single-conjugate adaptive optics (SCAO), which requires only two stars and does not rely on the assumption that tip-tilt varies linearly across the field, is to use tip-tilt tomography. Following the lead of Whiteley {\em et al},\cite{Whiteley,TokovininTomography} we use covariance matrices to produce the minimum variance estimate of the on-axis tip-tilt, $\hat a=[\hat a_x,\hat a_y]$ based on two or more off-axis tip-tilt measurements, $s$:

\begin{eqnarray}
\hat a &=&C_{as}C_{ss}^{-1}s \nonumber\\
&=&\mathit{Ms}
\end{eqnarray}
where  $s=[[s_{1x},s_{1y}],[s_{2x},s_{2y}],...,[s_{Nx},s_{Ny}]]$ are the $x$ and $y$ tip-tilt measurements for stars $1, 2, \dots$, $N$. Matrix $M$ is the reconstructor matrix, which converts the $2N$ off-axis measurements to on-axis tip-tilt. 

A very complicated expression for the calculation of covariance matrices is described by Whiteley {\em et al}.\cite{Whiteley} In this paper, we show a direct calculation of the covariance matrices using numerical integration. The integrand is based on the filter function theory developed by Sasiela, \cite{Sasiela} with results found to be identical to those obtained using Whiteley's formulation.\cite{ConanPersonalCommunication} We present it here as it is not published elsewhere.

Without loss of generality, we assume that two stars are located at at $[0,0]$ and $[\theta,0]$. Then
\begin{eqnarray}
\label{eq:covcalc}
\left\langle a_x(0,0)a_x(\theta ,0)\right\rangle &=&\left\langle a_x(0,0)a_x(0,0)\right\rangle -0.5\left\langle
(a_x(0,0)-a_x(\theta ,0))^2\right\rangle \\
\left\langle a_x(0,0)a_y(\theta ,0)\right\rangle &=& 0 \\
\left\langle a_y(0,0)a_x(\theta ,0)\right\rangle &=&0 \\
\left\langle a_y(0,0)a_y(\theta ,0)\right\rangle &=&\left\langle a_y(0,0)a_y(0,0)\right\rangle -0.5\left\langle
(a_y(0,0)-a_y(\theta ,0))^2\right\rangle 
\end{eqnarray}

The right hand side of the equations in Eq. (\ref{eq:covcalc}) have two terms. The first term is the tip (or tilt) variance. The other terms are the parallel and perpendicular contributions to the anisokinetic error:

\begin{equation}
\left\langle (a_x(0,0)-a_x(\theta ,0))^2\right\rangle =\int _0^{\infty }\mathit{dh}\int _0^{\infty }\mathit{dk}2\pi
k\frac{0.0097C_n^2(h)}{(k^2+1/L_0^2)^{11/6}}\left(\frac{4J_2(2\pi Rk)}{2\pi Rk}\right)^2\left(1+2J_1(2\pi k\theta
h)-2J_0(2\pi k\theta h)\right) \nonumber
\end{equation}
and
\begin{equation}
\left\langle (a_y(0,0)-a_y(\theta ,0))^2\right\rangle =\int _0^{\infty }\mathit{dh}\int _0^{\infty }\mathit{dk}2\pi
k\frac{0.0097C_n^2(h)}{(k^2+1/L_0^2)^{11/6}}\left(\frac{4J_2(2\pi Rk)}{2\pi Rk}\right)^2\left(1-2J_1(2\pi k\theta 
h)\right)
\end{equation}
For small to modest offsets, the variance of the parallel component is three times the variance of the perpendicular component. 

The three components two-dimensional integral written in polar coordinates are: 
\begin{itemize}
\item[--] the von Karman power spectrum, $\frac{0.0097C_n^2(h)}{(k^2+1/L_0^2)^{11/6}}$ 
\item[--] the filter function for tip-tilt, $\left(\frac{4J_2(2\pi Rk)}{2\pi Rk}\right)^2$ 
\item[--] the anisoplanatic contribution in the parallel and perpendicular directions to the offset.
\end{itemize}

The variance of the tip-tilt terms is obtained by removing the anisoplanatic filter function:
\begin{equation}
\left\langle a_x(0,0)a_x(0,0)\right\rangle =\left\langle a_y(0,0)a_y(0,0)\right\rangle =\frac 1 2\int _0^{\infty
}\mathit{dh}\int _0^{\infty }\mathit{dk}2\pi k\frac{0.0097C_n^2(h)}{(k^2+1/L_0^2)^{11/6}}\left(\frac{4J_2(2\pi
Rk)}{2\pi Rk}\right)^2,
\end{equation}
where the factor of half divides the total tip-tilt power into tip and tilt components.
$R$ is the radius of the telescope, and all the other symbols have their usual meaning.

For completeness, we state the total anisokinetic error here, which is merely the sum of the parallel and perpendicular components:
\begin{equation}
\sigma ^2=\int _0^{\infty }\mathit{dh}\int _0^{\infty }\mathit{dk}2\pi
k\frac{0.0097C_n^2(h)}{(k^2+1/L_0^2)^{11/6}}\left(\frac{4J_2(2\pi Rk)}{2\pi Rk}\right)^2\left(2-2J_0(2\pi k\theta
h)\right)
\end{equation}

To calculate the covariance when the star offsets are not aligned with either the tip or tilt offsets requires the coordinates to be transformed using a rotation matrix. Once we have computed these covariance matrices, it is straight-forward to compute the tomographic error:
\begin{equation}
\sigma ^2=\rm{trace}\left(C_{aa}-C_{as}C_{ss}^{-1}C_{as}^T\right),
\end{equation}
or, in the case of noisy measurements with noise covariance matrix,  $C_{nn}$:
\begin{equation}
\sigma ^2=\rm{trace}\left(C_{aa}-C_{as}(C_{ss}+C_{nn})^{-1}C_{as}^T\right).
\end{equation}

The minimum variance reconstructor leads to optimal open-loop estimates, and this reconstructor can be used in closed-loop. However, it is not clear how to optimally combine tip-tilt measurements for the general case where two or more stars with very different magnitudes are read at different frame rates. This topic is addressed by Correia {\em et al.} in the context of MCAO, where the plate scale modes also need to be measured, and they found the unintuitive result that using an optimized single-rate controller always outperformed a multi-rate controller.\cite{CorreiaJOSA,CorreiaSPIE} For LTAO, an optimized multi-rate controller must outperform a single-rate controller, since only on-axis tip-tilt must be estimated. Further work on multi-rate controllers for single-conjugate adaptive optics systems is required. To be useful for NIR TTSs, these controllers should consider that measurements using off-axis stars are also have a centroiding error that depends on the shape of the spot on the detector, not just on the number of photons detected.

\subsection{Focus sensing}
\label{sec:lift}
In addition to tip-tilt measurements, focus measurements at a lower bandwidth are needed to track the changes in the sodium altitude, and some systems also measure quasi-static aberrations induced by LGS elongation.\cite{LGSAOoverview} A number of solutions to avoid the need for second expensive NIR camera have been proposed or implemented: 

\begin{enumerate}[nolistsep]
\item Use a visible light focus sensor operating on the same star as a visible light TTS. This solution, implemented on GeMS and on Keck I, requires a sufficiently bright gude star in the visible, which partially negates the benefits of a NIR TTS. 
\item Use a $2\times2$ SH WFS to measure tip-tilt and focus, as implemented in IRLOS.
\item Use the pixel intensities in a well sampled spot to deduce low-order aberrations.
\end{enumerate}
The first two solutions clearly work, with obvious drawbacks. The third solution appears to provide a ``free lunch''. It has been proposed to add a low-order aberration to the beam, and then deduce focus and other low-order terms based on pixel intensities.\cite{LIFT,LIFT2016} For example, it is well known that if \SI{45}{\degree} astigmatism is added to the beam, then a defocus in one direction will lead to an image elongation at \SI{45}{\degree}, while a defocus in the opposite direction will elongate the image at \SI{-45}{\degree}. This technique has been ``tested'' in simulations and on test benches without atmospheric turbulence. However, there are multiple real-life effects which also elongate the image including wind-shake, anisoplanatic effect (for off-axis guide stars), so-called LGS aberrations,\cite{LGSAOaberrations} calibration errors and telescope phasing errors (for segmented telescopes). These risks need to be retired or mitigated. A possible solution would be to add dynamic rather than static aberrations. In addition, the use of LIFT requires the PSF to be sampled much more finely than what is required to estimate the spot displacement.

\section{Dicussion and conclusion}
\label{sec:conclusion}

This paper describes the implementation and performance of the three existing IR TTSs implemented on \SI{8}{}-\SI{10}{\meter} class telescopes. The use of IR TTSs has the potential to greatly increase the sky coverage of these LGS-based AO systems, and excellent results have already been demonstrated using relatively bright, near-on axis tip-tilt stars. The use of Hawaii-xRG detectors defines an envelope of frame rate, ROI size and read noise parameters which can be traded-off against each other, but ultimately restrict the limiting magnitude and the off-axis distance. New detectors now exist where the full frame can be read at kilohertz rates with subelectron read noise, and these should revolutionize the use of IR TTSs.

IR TTSs are also trickier to operate than conventional visible light TTSs because they require diffraction-limited PSFs, and hence full AO correction, in order to close the tip-tilt loop. Reacquisition is required if the high-order loop is lost.

There are still a number of unsolved research problems and these are listed here:
\begin{itemize}[nolistsep]
\item[--] Measuring centroids from partially corrected PSFs ({\em e.g.}, off-axis stars). 
\item[--] Measuring centroids from faint, extended sources.
\item[--] Optimal on-axis tip-tilt estimates from two or more tip-tilt measurements at different frame rates.
\item[--] Measuring focus and other low-order terms from PSF in real-life ({\em e.g.}, in the presence of wind-shake, off-axis elongation, quasi-static aberrations) .  
\end{itemize}

\section*{Acknowledgments}

The authors wish to recognize and acknowledge the very significant cultural role and reverence that the summit of Maunakea has always had within the indigenous Hawaiian community. We are most fortunate to have the opportunity to conduct observations from this mountain. 



\begin{thebibliography}{1}

\bibitem{LGSAOoverview}
P.~L. Wizinowich {\em et al.}, ``The WM Keck Observatory laser guide star adaptive optics system: overview,'' PASP {\bf 118}, 297 (2006).

\bibitem{SAPHIRA}
G. Finger {\em et al.}, ``Sub-electron read noise and millisecond full-frame readout with the near infrared eAPD array SAPHIRA,'' Proc. SPIE 9909 (2016).

\bibitem{NIRWFSWizinowich}
P. Wizinowich {\em et al.}, ``Near-infrared wavefront sensing,'' Proc. SPIE 9902, 990925 (2016).

\bibitem{GMTDFS}
D. Kopon {\em et al.}, ``Preliminary on-sky results of the next generation GMT phasing sensor prototype,'' Proc SPIE 10703 (2018).

\bibitem{GeMSI}
F. Rigaut {\em et al.}, ``Gemini multiconjugate adaptive optics system review–I. Design, trade-offs and integration,'' MNRAS {\bf 437}, 2361-2375 (2013).

\bibitem{GeMSII}
B. Neichel {\em et al.}, ``Gemini multiconjugate adaptive optics system review–II. Commissioning, operation and overall performance,'' MNRAS {\bf 440}, 1002-1019 (2014).

\bibitem{NGS2}
F. Rigaut {\em et al.}, ``NGS2: a focal plane array upgrade for the GeMS multiple tip-tilt wavefront sensor." Proc SPIE {\bf 9909}, (2016).

\bibitem{Young}
P.~J. Young {\em et al.}, ``Using ODGWs with GSAOI: software and firmware implementation challenges,'' Proc SPIE {\bf 8451}, 845124 (2012).

\bibitem{LGSAOperformance}
M.~A. van Dam {\em et al.,} ``The WM Keck Observatory laser guide star adaptive optics system: performance characterization,'' PASP {\bf 118}, 310 (2006).

\bibitem{KeckAOUpgrade}
E.~M. Johansson {\em et al.}, ``Upgrading the Keck AO wavefront controllers,'' Adaptive Optics Systems. Vol. 7015. International Society for Optics and Photonics (2008).

\bibitem{TRICK2014}
P. Wizinowich {\em et al.}, ``A near-infrared tip-tilt sensor for the Keck I laser guide star adaptive optics system,'' Proc. SPIE {\bf 9148} 91482B (2014).

\bibitem{TRICK2015}
R. Rampy {\em et al.}, ``Near-infrared tip-tilt sensing at Keck: System architecture and on-sky performance,'' AO4ELT4 (2015).

\bibitem{TRICK2016}
B. Femen\'ia Castell\'a {\em et al.}, ``Status and new developments with the Keck I near-infrared tip-tilt sensor,'' Proc. SPIE {\bf 9909}, 990925 (2016).

\bibitem{Smith}
R. Smith, and D. Hale, ``Read noise for a 2.5 micron cutoff Teledyne H2RG at 1-1000Hz frame rates,'' SPIE Proc. 8453, 84530Y (2012).

\bibitem{TRICKmultipleGS}
C. Samulski {\em et al.}, ``Optimizing use of multiple stars for near-infrared tip-tilt compensation at the WM Keck Observatory,'' AO4ELT4 (2015).

\bibitem{obertispie2018}
S. Oberti {\em et al.}, ``The AO in AOF,'' Proc. SPIE {\bf 10703}, 107031G (2018).

\bibitem{Gach}
J.-L. Gach {\em et al.}, ``Infrared detectors for wavefront sensing,'' AO4ELT5 149 (2017).

\bibitem{LIFT}
S. Meimon {\em et al.}, ``LIFT: a focal-plane wavefront sensor for real-time low-order sensing on faint sources,” Optics Letters {\bf 35}, 3036-3038 (2010).

\bibitem{LIFT2016}
C. Plantet {\em et al.}, ``LIFT on Keck: analysis of performance and first experimental results,'' Proc. SPIE 9909 (2016).

\bibitem{EnvironmentalTestingDMs}
I. Price and F. Rigaut, ``Environmental Testing of Deformable Mirrors: II Temperature Dependence,'' Australian National University Report GMTIFS-OIWFS-REP-0002, Version 1.2 (2017).

\bibitem{FlickerTTaniso}
R.~C. Flicker, F.~J. Rigaut, and B.~ L. Ellerbroek, ``Tilt anisoplanatism in laser-guide-star-based multiconjugate adaptive optics,'' A\&A {\bf 400}, 1199–1207 (2003).

\bibitem{Whiteley}
M.~R. Whiteley {\em et al.}, ``Optimal modal wave-front compensation for anisoplanatism in adaptive optics,'' JOSA A {\bf 15}, 2097-2106 (1998).

\bibitem{TokovininTomography}
A. Tokovinin {\em et al.}, ``Optimized modal tomography in adaptive optics,'' Astronomy \& Astrophysics {\bf 378}, 710-721 (2001).

\bibitem{Sasiela}
R. Sasiela, ``Electromagnetic wave propagation in turbulence,'' 2\textsuperscript{nd} edition, SPIE Press (2007).

\bibitem{ConanPersonalCommunication}
R. Conan, personal communication (15 March 2011).

\bibitem{CorreiaJOSA}
C. Correia {\em  et al.}, ``Increased sky coverage with optimal correction of tilt and tilt-anisoplanatism modes in laser-guide-star multiconjugate adaptive optics,'' JOSA A {\bf 30}, 604-615 (2013).

\bibitem{CorreiaSPIE}
C. Correia {\em et al.}, ``Advanced control of low order modes in laser guide star multi-conjugate adaptive optics systems'' Proc. SPIE {\bf 8447}, 84471S (2012).

\bibitem{LGSAOaberrations}
R.~M. Clare, M.~A. van Dam, and A.~H. Bouchez, ``Modeling low order aberrations in laser guide star adaptive optics systems,'' Optics Express {\bf 15}, 4711-4725 (2007).

\end{thebibliography}
\end{document}